# Stoichiometric lithium niobate crystals: towards identifiable wireless surface acoustic wave sensors operable up to 600°C

Jérémy Streque[3], Thierry Aubert[1,2], Ninel Kokanyan[1,2], Florian Bartoli[1,2,3]*, Amine Taguett[1,2], Vincent Polewczyk[3], Edvard Kokanyan[4], Sami Hage-Ali[3]*, Pascal Boulet[3] and Omar Elmazria[3]*

[1] *Laboratoire Matériaux Optiques, Photonique et Systèmes (LMOPS), CentraleSupélec, Université Paris-Saclay, Metz, F-57070, France*
[2] *Université de Lorraine, Matériaux Optiques, Photonique et Systemes (LMOPS), Metz, F-57070, France*
[3] *Institut Jean Lamour, UMR 7198 Université de Lorraine – CNRS, 2 Allée André Guinier, site Artem, 54000 Nancy, France*
[4] *Institute for Physical Research, National Academy of Sciences of Armenia, 0203 Ashtarak-2, Armenia and the Armenian State Pedagogical University after Kh. Abovyan, 0010 Yerevan, T. Mets av. 17, Armenia.*
* Member, IEEE

**Abstract—** Wireless surface acoustic wave (SAW) sensors constitute a promising solution to some unsolved industrial sensing issues taking place at high temperatures. Currently, this technology enables wireless measurements up to 600-700°C at best. However, the applicability of such sensors remains incomplete since they do not allow identification above 400°C. The latter would require the use of a piezoelectric substrate providing a large electromechanical coupling coefficient $K^2$, while being stable at high temperature. In this letter, we investigate the potentiality of stoichiometric lithium niobate (sLN) crystals for such purpose. Raman spectroscopy and X-ray diffraction attest that sLN crystals withstand high temperatures up to 800°C, at least for several days. *In situ* measurements of sLN-based SAW resonators conducted up to 600°C show that the $K^2$ of these crystals remains high and stable throughout the whole experiment, which is very promising for the future achievement of identifiable wireless high-temperature SAW sensors.

## I. INTRODUCTION

SAW devices are key components of mobile phone industry and other telecommunication systems, but are also very promising for sensing various physical quantities. They offer exciting perspectives for remote monitoring and control of moving parts, especially in harsh environments. Indeed, SAW sensors are passive devices, and just re-radiate a small part of the energy received from the RF interrogation signal [1, 2].

Wireless SAW sensors capable of measuring temperatures up to 600-700°C already exist [3-5] and are based on langasite crystals ($La_3Ga_5SiO_{14}$, often called LGS). The surface of these crystals shows an outstanding stability at high temperatures up to 1000°C in air atmosphere [6]. Due to relatively low electromechanical coupling coefficient $K^2$ values (typically 0.4%), LGS-based SAW sensors are configured as resonators, whose identification can only be performed by frequency segmentation, and thus for a few sensor units simultaneously. This limitation can be overtaken by the use of the reflective delay line (R-DL) configuration, making the sensors identifiable through their coded time-resolved response [7]. High-performance R-DL sensors require time pulses as short as possible, and thus the largest possible device bandwidth. In this respect, with typical $K^2$ values up to 5% and a Curie temperature above 1000°C, ferroelectric lithium niobate ($LiNbO_3$, often called LN) crystals are naturally strong candidates to achieve high temperature R-DL SAW sensors. LN crystals span from quasi-stoichiometric compositions to lithium-poor grades as low as approximately 45 mol% $Li_2O$ [8]. Most of commercially available high-quality LN single crystals are produced by Czochralski technique and have a composition near the congruently melting value of roughly 48.4 mol% $Li_2O$. Single crystals of stoichiometric composition, grown either by the top-seeded solution growth (TSSG) method from potassium-containing flux [9] or from lithium-rich melts by double crucible method [10], are also commercially available, but in smaller quantities.

SAW devices based on congruent LN (cLN) were studied at high temperatures over a decade ago. They are limited to temperatures below 400°C for at least two different reasons. First, cLN crystals are not thermodynamically stable below 900°C. The segregation of a niobium-rich phase $LiNb_3O_8$ occurs in such conditions [11-12]. The kinetics of this process becomes really problematic above 400°C [13]. Moreover, the time pulses attenuation of cLN-based R-DL increases drastically with temperature, prohibiting such devices to be remotely interrogated above 400°C [14]. Such effect could be related to the relatively low electrical resistivity of cLN crystals at high temperature due to Li vacancies [15-16].

Corresponding author: T. Aubert (thierry.aubert@centralesupelec.fr).

On the contrary, stoichiometric lithium niobate (sLN) crystals are expected to be thermodynamically stable up to their melting temperature at 1170°C [11], while keeping a larger electrical resistivity than cLN crystals by one order of magnitude at any temperature [17]. Promising bulk acoustic wave (BAW) signals were observed up to 880°C on sLN-based BAW resonators, with a stable quality factor Q up to 700°C [18-19]. The purpose of the present letter is to explore the potential of sLN crystals for high temperature SAW applications, in view of the future achievement of identifiable wireless R-DL sensors able to operate in the intermediate temperature range, i.e. from 400 to 600°C at least.

## II. EXPERIMENTAL SETUPS

Piezoelectric substrates need to fulfill various requirements for high temperature SAW sensing compatibility: while structural and chemical stability are mostly desirable for recovering from multiple temperature variations over the lifetime of the devices, the electroacoustic properties must be high enough over the whole operating temperature range. Consequently, material characterizations were firstly performed in order to assess the sLN stability with temperature, in terms of structure and composition. Then the electroacoustic properties were examined through the study of a first series of SAW devices fabricated to this purpose.

The investigated samples consist in Z-cut near-stoichiometric (49.8 mol% $Li_2O$) LN substrates (MTI Corp., Richmond, CA). Some sLN crystals were placed in a furnace to be heated for 48h at 800°C in air. The evolution of the crystallographic structure was analyzed by X-ray diffraction (XRD), in Bragg-Brentano geometry (PANalytical X'Pert Pro MRD - CuKα1: $\lambda$ = 1.54056 Å). Additional information regarding this evolution was provided by Raman spectroscopy measurements (Horiba Jobin Yvon ARAMIS). A He-Ne laser with 632.8 nm of wavelength was operated. The penetration depth was about 1 μm. Complementary to XRD measurements, Raman spectroscopy gives the possibility to reveal the formation of a thin amorphous surface layer. It also gives access to the stoichiometry of the crystal surface. To do so, the spectra of Z(XY)Z backscattering configuration were measured in order to obtain the E[$TO_1$] Raman mode, corresponding to the Nb/O vibration in the (X,Y) plane [20], which provides a possibility to calculate the composition of the crystal from the linewidth of this mode [21]. The Raman and XRD results were compared with those obtained from similar experiments conducted on Z-cut cLN crystals (48.4 mol% $Li_2O$) purchased at the same supplier. The remaining samples were then employed to investigate some electromechanical properties of sLN crystals at high temperature, in particular $K^2$ as it is an essential parameter for the achievement of R-DL sensors. These electromechanical parameters were accessed by a classical method, based on the measurement of SAW resonators. It is worth noting that the investigated resonators are not envisioned as sensors but only as a probe to measure these parameters.

Regarding the fabrication of the resonators, 150 nm-thick aluminium films were deposited by DC magnetron sputtering on the sLN substrates for the patterning of the electrodes. Aluminum is not the most suitable material to achieve electrodes able to sustain high-temperature conditions. However, it can withstand temperatures up to 600°C for short time characterizations like those conducted in this study (typically three hours for each heating-cooling cycle) and its remarkable properties in term of low density and electrical conductivity makes it particularly convenient for the design of efficient SAW resonators.

Synchronous single-port resonators with a wavelength of $\lambda$ = 6.5 μm were fabricated by conventional photolithography and wet etching. They were equipped with 200 reflectors on each side of their InterDigitated Transducer (IDT), which was constituted by 100 finger pairs, with an aperture of 40·$\lambda$. The metallization ratio of the electrodes was set to 40%. The direction of wave propagation was along the Y-direction of the crystals. These devices were *in situ* electrically characterized under air atmosphere, using a network analyzer and an RF prober station (S-1160, Signatone Corp., Gilroy, CA) equipped with a thermal probing system that enables to control the device temperature up to 600°C (S-1060, Signatone).

## III. RESULTS AND DISCUSSION

Raman measurements conducted before annealing exactly confirmed the near-stoichiometric composition of the investigated crystals, with a measured lithium concentration of 49.85±0.05 mol% $Li_2O$, in perfect agreement with the manufacturer data. No measurable change in the whole Raman spectra and in the E[$TO_1$] Raman mode in particular could be observed after the 48h-long annealing treatment at 800°C (Fig. 1). These results indicate that the crystal surface does not undergo any lithium losses, keeping the stoichiometric composition and the initial microstructure.

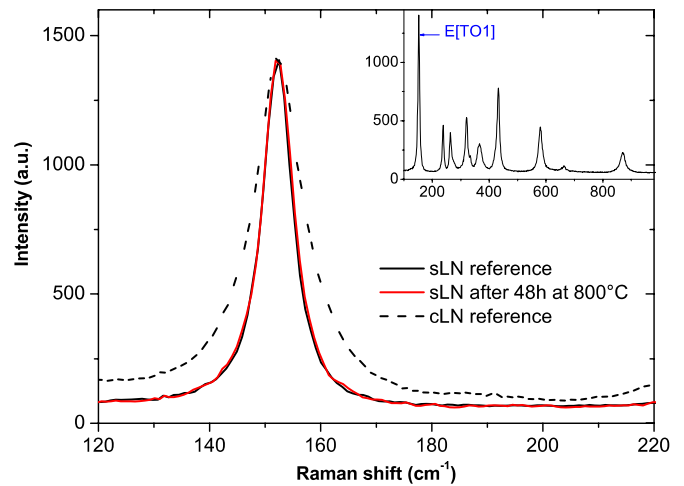

Fig. 1. E[$TO_1$] Raman mode of sLN crystals before and after a 48h-long annealing process at 800°C. The same mode obtained for cLN crystal was added for comparison. In the inset can be seen the whole Raman spectra of the sLN crystal. All spectra were carried out in Z(XY)Z configuration.

XRD measurements confirm that the lattice structure of the sLN crystals is unchanged after this annealing process (Fig. 2(a)), in contrast with cLN crystals. In that case, the (006) $LiNbO_3$ reflex is shifted towards low angles by 0.07°, while its intensity is divided by a factor 3 after the heating process. Moreover, a new broad peak appears at 2θ = 37.85°, which can be identified as the (60-2) LiNb$_3$O$_8$ reflex (Fig. 2(b)), confirming that cLN crystals segregate in such conditions contrariwise to sLN crystals.

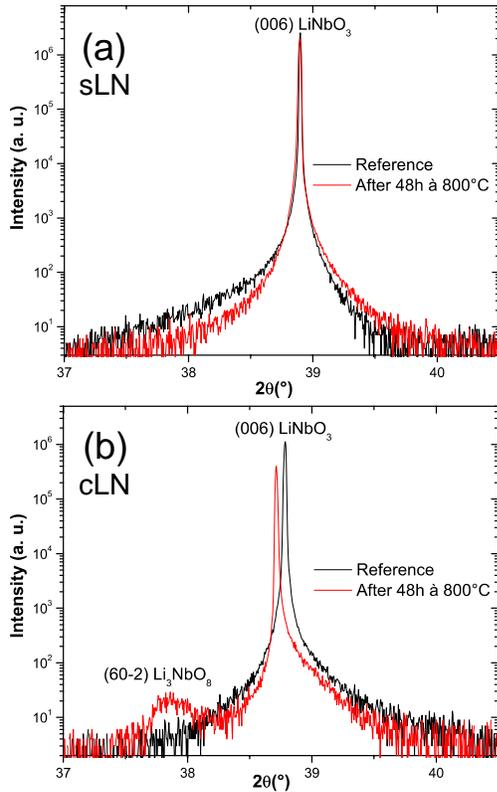

Fig. 2. *θ-2θ* XRD patterns of reference and 48h-annealed at 800°C sLN (a) and cLN (b) crystals.

Based on these structural characterizations, it can be assumed that sLN crystals can be employed as piezoelectric substrates in SAW sensors operating up to 600°C for long periods of at least several days. Their ability to preserve good electroacoustic properties have then been verified through the RF characterization of the fabricated devices between room temperature and 600°C, as well as during the cooling phase. The measured admittance spectra are shown in Figure 3. The operating frequency is close to 603 MHz at room temperature, which corresponds to a SAW velocity of 3920 m/s, in very good agreement with the literature regarding the propagation of a Rayleigh wave on ZY-cut LN crystals [22].

The sLN crystals deliver a linear response to temperature variations, and the frequency-temperature law is reproducible which supports the fact that the aluminium electrodes withstand this characterization process (Fig. 4). The sensitivity of the crystal to temperature variations is high with a temperature coefficient of frequency (TCF) of -84 ppm/K, which is appropriate for the achievement of temperature sensors.

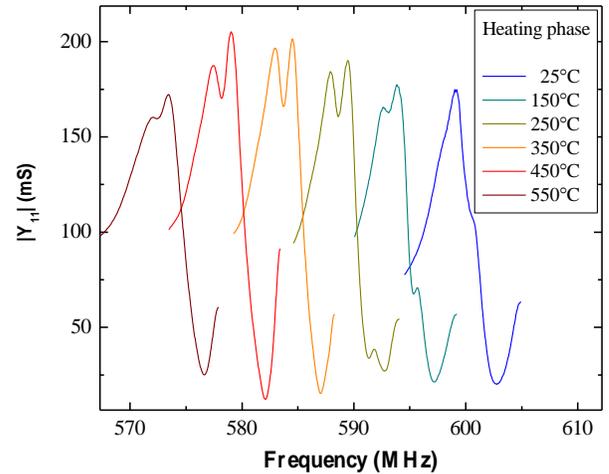

Fig. 3. Admittance magnitude of an sLN-based resonator measured at different temperatures during the heating phase.

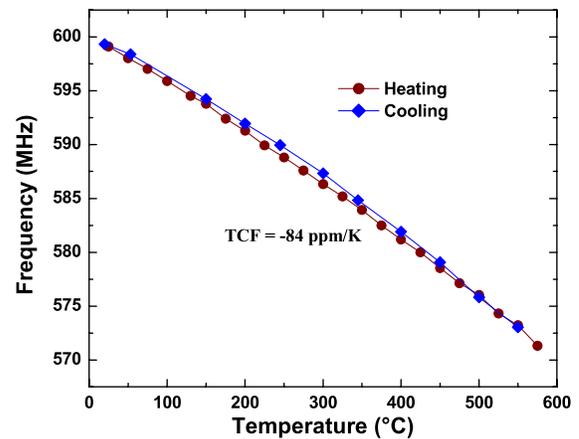

Fig. 4. Frequency-temperature law of the sLN-based SAW resonators measured between the ambient and 600°C, during the heating and the cooling phases.

Effective electromechanical coupling coefficient values have then been derived from the admittance measurements, through the determination of the parallel and series resonance frequencies, $f_p$ and $f_s$ [23]:

$$K_{eff}^2 = \frac{\pi^2}{4} \frac{f_p - f_s}{f_p}$$

$K_{eff}^2$ remains stable, with values close to 1.4% ± 0.2%, in the whole investigated temperature range (Fig. 5). The small observed variations of $K_{eff}^2$ are mainly attributed to the uncertainties related to the reading of the parallel and series resonance frequencies on the admittance curves, due to the presence of small spurious modes, as these variations are not completely repeatable from a device to the other.

Beyond these small variations, no significant drop of $K^2$ can be observed over the whole temperature range of interest, which is very promising for the future achievement of identifiable high-temperature R-DL SAW sensors.

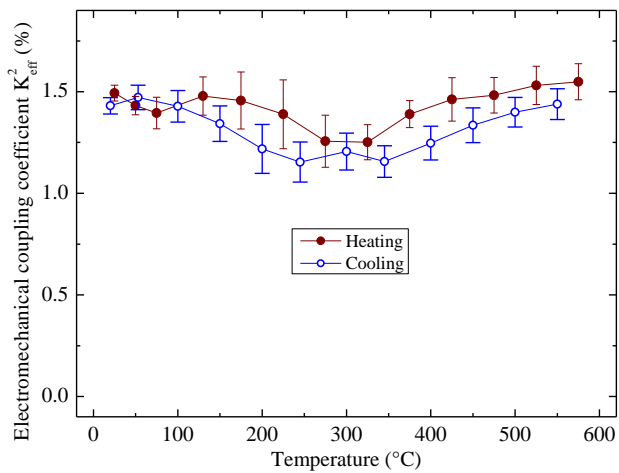

Fig. 5. Stability in temperature of the effective $K^2$ coefficient calculated from admittance measurements on sLN-based resonators.

## IV. CONCLUSION

sLN crystals have been investigated in the purpose of being used for high-temperature SAW applications in air atmosphere. Raman and XRD measurements attest that the crystals are not affected by a 48h-long heating process at 800°C in air, in contrast with LN crystals of congruent composition. *In situ* SAW measurements conducted on sLN-based resonators between room temperature and 600°C confirm the suitability of sLN crystals for high-temperature SAW applications, in particular for sensing temperature as the sensitivity of these crystals to this parameter is high and linear. The electromechanical coupling coefficient retrieved from RF measurements is very stable, in the order of 1.5% in the whole investigated temperature range. These promising results pave the way to the realization of high-temperature R-DL sensors. However, many challenges will have to be faced before this goal can be achieved. In particular, it will be necessary to study the behavior of SAW propagation losses on sLN crystals at high temperatures. Moreover, it will be mandatory to introduce low-density and low–resistivity electrodes offering good thermal stability, allowing the generation of strong SAW pulses in the 2.45 GHz frequency band, whose large bandwidth makes it the most suited ISM band for the design of R-DL sensors.